\documentclass[num-refs]{wiley-article}

\usepackage{siunitx}
\usepackage[english]{babel}
\usepackage{tabulary}
\usepackage{xspace}
\usepackage{dcolumn}
\usepackage{bm}
\usepackage{booktabs}
\usepackage{tikz,pgf,epsfig,wasysym,mathrsfs}
\usepackage{xcolor}
\usepackage{siunitx}
\usepackage{multirow}
\usepackage{amsmath}
\usepackage{braket}
\usepackage{mathtools}

\usepackage[ruled,linesnumbered]{algorithm2e}

\newcommand{\I}{\mathcal{I}}
\newtheorem{fact}{Fact}
\ifx\newdefinition\undefined
  \let\newdefinition\newtheorem
\fi
\ifx\definition\undefined
  \newdefinition{definition}{Definition}

\fi

\papertype{Original Article}
\paperfield{Journal Section}

\title{A Hybrid Quantum-assisted Column Generation Algorithm for the Fleet
Conversion Problem}

\author[1,2]{Yagnik Chatterjee}
\author[1]{Zaid Allybokus}
\author[1\authfn{2}]{Marko J. Ran\v{c}i\'{c}}
\author[2]{Eric Bourreau}

\affil[1]{TotalEnergies, Tour Coupole - 2 place Jean Millier 92078 Paris la Défense cedex, France}
\affil[2]{LIRMM, Université de Montpellier, CNRS, 161 Rue Ada, Montpellier, France}

\corraddress{Yagnik Chatterjee, TotalEnergies, Tour Coupole - 2 place Jean Millier 92078 Paris la Défense cedex, France}
\corremail{yagnik.chatterjee@totalenergies.com}

\presentadd[\authfn{2}]{University of Heidelberg, Institute of Computer Engineering, Im Neuenheimer Feld 368, 69120 Heidelberg, Germany}

\fundinginfo{European Union Horizon 2020, $\bra{\text{NE}}\text{AS}\ket{\text{QC}}$ project, Grant Number: 951821}

\runningauthor{Chatterjee et al.}

\begin{document}

\begin{frontmatter}
\maketitle

\begin{abstract}
The problem of Fleet Conversion aims to reduce the carbon emissions and cost of operating a fleet of vehicles for a given set of tours. It can be modelled as a column generation scheme with the Maximum Weighted Independent Set(MWIS) problem as the sub-problem or worker problem. Quantum variational algorithms have gained significant interest in the past several years. Recently, a method to represent Quadratic Unconstrained Binary Optimization(QUBO) problems using logarithmically fewer qubits was proposed. Here we use this method to solve the MWIS Workers and demonstrate how quantum and classical solvers can be used together to approach an industrial-sized use-case (up to 64 tours).

\end{abstract}
\end{frontmatter}

\section{Introduction}\label{sec:1}

Fleet conversion is the process of transitioning a fleet of vehicles to more sustainable and environmentally friendly alternatives. With the growing recognition of the detrimental effects of traditional fossil fuel powered vehicles on the environment and the need to mitigate climate change, businesses and organizations are increasingly looking for ways to reduce their carbon footprint and operate more efficiently. The transportation sector is one of the largest contributors to greenhouse gas emissions, primarily due to their reliance on fossil fuels. By transitioning fleets to electric or hybrid vehicles, companies can significantly reduce their carbon emissions. Beyond the environmental benefits, fleet conversion also offers compelling cost-saving opportunities for businesses. 

In the fleet conversion problem, a certain number of tours need to be carried out between several locations. In order to carry out these tours we have at our disposal several vehicles of different models. Each vehicle model has an associated cost. On top of the capital expenditure corresponding to the purchase of one vehicle of one model, this cost may also capture the environmental cost -- e.g. the carbon footprint; the cost of operation -- e.g. energy usage, or both. The objective is to minimize the total cost of carrying out all the tours including capital and operational expenditures. Therefore, fleet conversion goes beyond simply choosing the best possible vehicles and also incorporates sharing the same vehicles for multiple tours when possible, thereby reducing the cost.

Quantum computing \cite{neilsen,Preskill2021,Preskill2018quantumcomputingin} is a potentially disruptive field that could have applications in several domains including financial modelling \cite{kitaev1995quantum,herman2022survey}, cryptography \cite{scarani2009security,mehic2020quantum}, chemistry \cite{martin2022simulating,haidar2023extension} and optimization.  Within the scope of optimization applications, there has been a growing interest in quantum variational algorithms \cite{cerezo2021variational,peruzzo2014variational,moll2018quantum,lubasch2020variational,stokes2020quantum}. Among them, the Quantum Approximate Optimization Algorithm (QAOA) has been heavily researched \cite{qaoa, zhou2020quantum,qaoa2}. A well known issue with QAOA is that it does not scale well with problem size limiting its applications to toy problems. Recently, an algorithm to treat Quadratic Unconstrained Binary Optimization (QUBO) \cite{qubotutorial,papalitsas2019qubo} problem using \textit{logarithmically} fewer qubits has been demonstrated\cite{rancic,chatterjee2023solving}. In this paper, we use column generation \cite{desaulniers2006column,mehrotra1996column} to describe our problem as a coordinator problem and several sub-problems henceforth referred to as workers. The worker problem in our case is the Maximum Weighted Independent Set (MWIS) problem which can be represented as a QUBO problem. We propose an algorithm that handles the coordinator problem using a commercial linear program solver Gurobi and the worker problems using a quantum solver based on \cite{chatterjee2023solving}. In our experiments, we solve instances up to a size of 64 tours using only 7 qubits to represent the MWIS workers. This shows that the method is compatible with the quantum computers of the Noisy Intermediate Scale Quantum (NISQ) era, where the quantum computers are still relatively small and are not free of noise.

The paper is structured as follows. In section \ref{Statement} the fleet conversion problem is defined. In section \ref{GraphDefinition} the problem is stated in the form of a graph problem followed by section \ref{ColumnGen} where the column generation algorithm is described. In section \ref{QuantumModel} and \ref{QuantumAlg}, we describe the quantum model to solve the sub-problems and how we can use the quantum solver and classical solver together to develop a quantum-assisted algorithm. Finally, we present the experimental results in section \ref{Experiments}.

\section{Problem Statement and Methods}\label{FleetConversion}

In this section, we present the Fleet Conversion Problem as well as its formulation as a weighted graph coloring problem. We then reformulate the problem using the definition of independent sets and build a column generation approach to solve it. The column generation approach uses sub-problems that compute max-weighted independent sets, further brought together iteratively to build a global graph coloring solution. We then demonstrate how a quantum algorithm can be crafted to solve these sub-problems and integrate the column generation procedure.

\subsection{Statement}\label{Statement}

Let $L$ be a set of locations, $K$ a set of tours, $V$ a set of available vehicle models, and $C$ a set of physical vehicles, henceforth referred to as \textit{color}.

\noindent \textbf{Notations.} 

\noindent For any physical vehicle $c$ of a certain model $v$, let us write $v(c) := v$.
Tours are carried out from one location to another. Let $w^k_v$ be the cost to assign a vehicle model $v$ to any tour $k$. Assigning a model $v$ to a tour $k$ means that the tour $k$ has to be carried out by a physical vehicle (color) from model $v$. To do so, some color $c$ such that $v(c) = v$ has to be assigned to tour $k$. The cost $w_v^k$ captures the operational expenditures incurred by performing tour $k$ with a vehicle model $v$. Since every color belongs a specific vehicle model, let the cost to assign the physical vehicle (color) $c$ to any tour $k$ be $\Gamma^k_{c}\coloneqq w^k_v$ when $v = v(c)$. We also define a cost $\gamma_{v(c)} > 0$ for using a color $c$ at least once. Clearly,  $\gamma_v$ represents the cost of purchase of one physical vehicle of model $v$ and we reasonably assume this cost is independent of the color (all vehicles of the same model are equivalent). Thus, without risk of confusion, we define $\gamma_c \coloneqq \gamma_{v(c)}$.

Each tour $k\in K$ is described by the tuple $(t_d^k,t_a^k,l_d^k,l_a^k, A^k)$, corresponding respectively to the departure time, arrival time, departure location and arrival location, and a set of authorized vehicle models of the tour. The time to travel from location $i$ to location $j$ ($TT_{ij}$)  can be computed using the distance matrix of the locations. This matrix is used to derive the compatibility time needed to relocate any physical vehicle from the arrival location of a tour to the departure location of another tour in case these tours are meant to be assigned to the same color.

\noindent \textbf{Incompatibilities.}

\noindent In the Fleet Conversion Problem, we have two types of incompatibilities:
\begin{itemize}
    \item \emph{tour-color incompatibilities}: a tour $k$ cannot be assigned to a color from a certain model $v$. In practice, this incompatibility can model constraints in urban mobility such as the forbidden penetration of internal combustion engines into low emission zones or simply drivers' personal preferences (manual \emph{vs} automatic, plug-in hybrid \emph{vs} electric, \emph{etc}). Let $A^k$ be the set of authorized colors for tour $k$. Specifically, a color $c$ can be assigned to tour $k\in K$ only if $v(c)$ is in the set of allowed models $A^k$ for that tour.
    \item \emph{tour-tour incompatibilities}: two different tours cannot share the same color (physical vehicle) because they occur at the same time, or because their departure and/or arrival locations make it impossible to transition in an acceptable time without perturbing the global schedule. Let $\I \subset \binom{{K}}{2}$
    be the set of unordered couples with a tour-tour incompatibility\footnote{For any set $X$ and any integer $n$,  $\binom{X}{n} \coloneqq \{I \subset X | \mbox{ }|I| = n \}$. }.
\end{itemize}

\noindent \textbf{First formulation.} 

\noindent The aim of the Fleet Conversion Problem is to minimize the overall cost $\sum_c \gamma_{c} y_c + \sum_{k,c} \Gamma^k_cx_c^k$, where

\begin{equation}\label{xallotcolor}
    x^k_c =\left\{ \begin{matrix}
        1 & \mbox{if color $c$ is assigned to tour $k$}\\
        0 & \mbox{otherwise,}
    \end{matrix}\right.
\end{equation}

and
\begin{equation}
    y_c =\left\{ \begin{matrix}
        1 & \mbox{if color $c$ is purchased}\\
        0 & \mbox{otherwise.}
    \end{matrix}\right.
\end{equation}
This can be achieved by preferring colors $c$ with low values of $\gamma_c$ and also by assigning multiple compatible tours to a single color while choosing the minimal values of $\Gamma_c^k$ if possible. This is to be done in such a way that all tours within $K$ are assigned to a compatible color.

Therefore, the Fleet Conversion Problem can be expressed as the following optimization problem:

\begin{align}
    \min & \sum_{c \in C} \gamma_c y_c + \sum_{k \in K, c\in C} \Gamma^k_cx_c^k & \label{eq:objective}\\
    \mbox{s.t. }    & \sum_c x^k_c \geq 1 & (\forall k\in K) \label{eq:assignment}\\
                   & x^k_c \leq y_c & (\forall c\in C)(\forall k\in K)\label{eq:opening}\\
                   & x^k_c + x^{k'}_c \leq 1 & (\forall \{k,k'\} \in \I)(\forall c\in C) \label{eq:tour-tour}\\
                   & x^k_c = 0 & (\forall k\in K)(\forall c\in C)(v(c) \notin A^k\label{eq:tour-color})\\
                   & y_c, x_c^k \in\{0,1\}. \label{eq:integrity}
\end{align}

Equation~\eqref{eq:assignment} states that a tour $k$ has to be assigned to at least one\footnote{In fact, \emph{exactly one} would be more appropriate than \emph{at least one}. However, as the variables $x_c^k$ are penalized by a cost in the objective, the two formulations are actually equivalent and at optimum, this constraint is actually active} color, whereas Equation~\eqref{eq:opening} forbids the assignment of tours to a color unless the color is purchased. Equation~\eqref{eq:tour-tour} forbids incompatible tours to share the same color and Equation~\eqref{eq:tour-color} avoids tour-color incompatibilities.

Problem~\eqref{eq:objective}--\eqref{eq:integrity} is thus a Mixed Integer Linear Program (MILP) that can be solved with for instance a commercial solver. However, as it is notoriously NP-complete as from $|C| \geq 3$, solver time performance will worsen quickly with problem size. For this reason, we reformulate Problem~\eqref{eq:objective}--\eqref{eq:integrity} and design an algorithm that will scale better.

\subsection{Formal description as a Graph Problem}\label{GraphDefinition}

Let $G=(K,E)$ be a graph where the nodes of the graph are the tours and the edges $(k,k')\in E$ of the graph denote the \textit{incompatibility} of the tours $k$ and $k'$. With the notations of the last paragraph, this means that we let $E := \mathcal{I}$. Two tours $i$ and $j$ are compatible if their time-windows do not overlap and we have enough time to travel from the arrival location of tour $i$ to the departure location of tour $j$, without loss of generality if tour $i$ occurs before tour $j$. Formally, we define $E=\{(i,j)| \forall i,j\in K,t_a^i+TT_{ij}>t_d^j\}$. For every tour $k$, we have a list of allowed models $A^k$.

\begin{figure}[htb!]
\centering
  \includegraphics[width=0.9\linewidth]{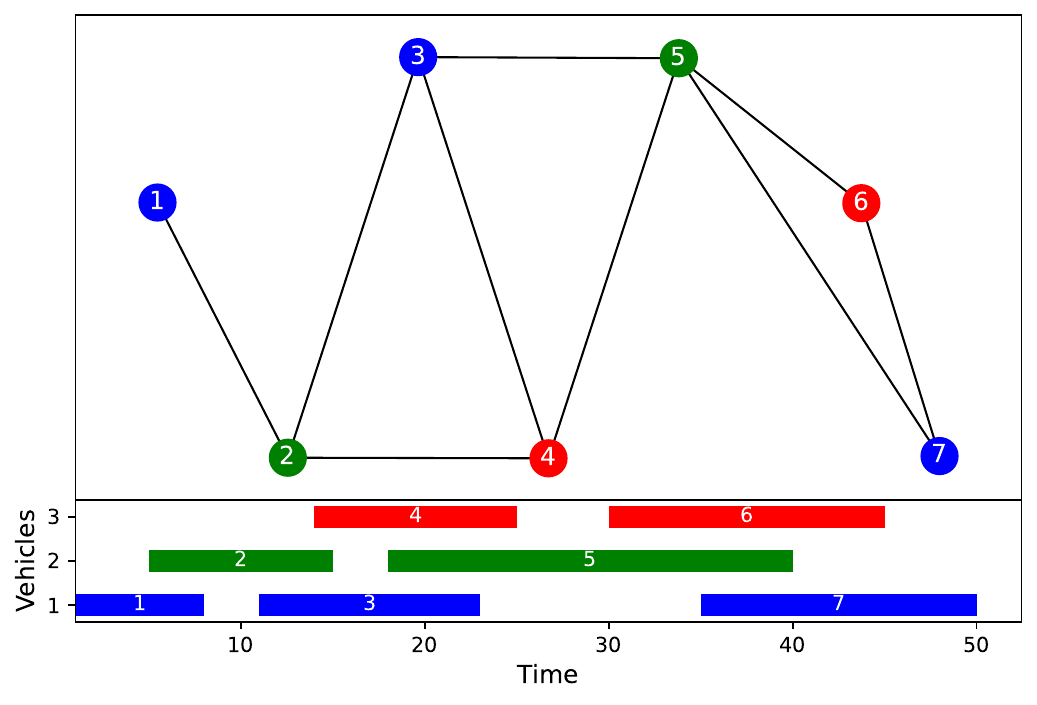}
  \caption{Incompatibility Graph generated from the time windows of tours. The lines represent the time windows of 7 tours. Colors \textit{blue}, \textit{red} and \textit{green} are used to assign vehicles to tours taking into account their incompatibility.}
  \label{interval}
\end{figure}

Figure \ref{interval} illustrates how an incompatibility graph can be constructed using time-windows of several tours, along with a possible coloring of the graph. For example, since the tour 3 overlaps with tours 2, 4 and 5, the node 3 has edges with node 2, 4 and 5.

\begin{definition}[Independent set]
    An independent set in $G = (K,E)$ is a subset $I \subseteq K$, such that for any two vertices $k, k' \in I$, $(k,k') \notin E$. In other words, an independent set of our graph is a subset of tours that are all compatible with each other to share colors. 
\end{definition}

With the definition of an independent set, the following fact is straightforward.

\begin{fact}
    Let $(x,y)$ be any solution of Problem~\eqref{eq:objective}--\eqref{eq:integrity}. Then, for all $c \in C$ such that $y_c = 1$, $\{ k | x_c^k = 1 \}$ forms an independent set of $G$.
\end{fact}

An independent set $I$ can be derived into an \emph{allocation} $(a_I,c_I)$, where $c_I\in C$ is a color and $a_I$ is the indicator vector of set $I$: $a^k_I=1$, if $k \in I$, and $0$ otherwise. For any independent set $I$ and color $c$, let $\Gamma_{I,c}:= \gamma_c + \sum_{k\in I}\Gamma^k_{c}$ denote the cost of allocation $(a_I,c)$. 
Here, we consider only  \emph{feasible} allocations, specifically respecting tour-color compatibility.
We can therefore define the set of feasible allocations $\Lambda$ formally as :
\begin{equation}
    \Lambda = \{(a_I,c_I)| I\mbox{ independent set of }G|\mbox{ s.t. }\forall k\in K, a^k_I = 1 \Rightarrow v(c_I)\in A^k\}.
\end{equation}

For lighter notations, we refer to the element  $(a_I,c_I) \in \Lambda$ only as $I \in \Lambda$, keeping in mind that the independent set $I$, considered as an allocation, comes with a color $c_I$. In particular, we keep in mind that if $c\neq c'$, and even if $v(c) = v(c')$, $(a_I, c)$ and $(a_I,c')$ are two different allocations. Formally, let $\Gamma_I \coloneqq \Gamma_{I, c_I}$.

The problem can then be equivalently formulated as:

\begin{align}
    \min & \sum_{I \in \Lambda} \Gamma_{I} x_{I}   & \label{eq:objective2}\\
    \mbox{s.t. }    & \sum_{I \in \Lambda} a_I^kx_I \geq 1 & (\forall k\in K) \label{eq:assignment2}
\end{align}

\begin{equation}\label{xindset}
  x_I = \begin{cases}
        1 , \text{ if independent set }I\text{ chosen }
        \\
        0 ,\text{ otherwise}
        \end{cases}
 \end{equation}


Note that constraint~\eqref{eq:assignment2} is equivalent to the constraint~\eqref{eq:assignment}. Also, beware that the variables defined in equation \eqref{xindset} and equation \eqref{xallotcolor} are not the same.

With the above reformulation, the problem looks considerably smaller. We have indeed only one constraint per tour. However, we need to generate the set $\Lambda$ in order to solve this problem. The set  $\Lambda$ contains all the feasible independent sets of the graph, that is, independent set of the graph coupled with colors that are compatible with all nodes therein. Remember that if  $I$ is a non-empty independent set of $G$ and $c \neq c'$ are two different colors, then $(I,c)$ and $(I,c')$ are two distinct elements of $\Lambda$ as they describe two different conversion solutions. 
We thus have a very large number of variables $x_I$. Nevertheless, it is clear that only a small number of those variables should be non-zero at optimum. Indeed, at worst (in terms of number of independent sets activated), no sharing of color is possible and each tour has a dedicated vehicle, which corresponds to $|K|$ variables $x_I$ activated. This is why an interesting approach here is column generation \cite{desaulniers2006column}, where independent sets are generated dynamically while the solution converges to an optimum.

\subsection{The Column Generation Algorithm}\label{ColumnGen}

In this paragraph, we design the column generation procedure to solve our problem. First, we derive an extension of the problem that makes it always feasible. To do so, we permit the algorithm to reject tours from the solution. Second, we relax all integrity constraints and actually solve the LP-relaxation of our problem. Once an LP-optimal solution is found by the column generation, one can summon any type of rounding algorithm to build (if needed) an integer-feasible solution from the relaxed solution. For instance, see \cite{bertsimas1999dependent}. We focus this work on the problem of finding the optimal LP-feasible solution for our problem. 

\vspace{5mm}
\noindent\textbf{Problem extension}
\vspace{5mm}

For all $k\in K$, let $r_k$ be a binary variable that states whether tour $k$ is rejected from the solution or not: 

\begin{equation}
    r_k = \left\{ \begin{matrix}
        1 & \mbox{if } k \mbox{ is rejected}\\
        0 & \mbox{otherwise.}
    \end{matrix}\right.
\end{equation}

Let $R$ be a sufficiently large real number.
We now consider the new optimization problem:

\begin{align} \tag{CP}
    \min & \sum_{I \in \Lambda} \Gamma_{I} x_{I} +R\sum_{k\in K} r_k   & \label{eq:objective3}\\
    \mbox{s.t. }    & \sum_{I \in \Lambda} a_I^kx_I + r_k \geq 1 & (\forall k\in K) \tag{CP-1}\label{eq:assignment3}\\
    & x_I, r_k \geq 0 & (\forall I\in \Lambda)(\forall k\in K)\tag{CP-2}\label{eq:positivity}
\end{align}

When the binary constraints are enforced on $x_I$ and $r_k$, and if $R$ is sufficiently large, it is clear that Problems~\eqref{eq:objective3} and~\eqref{eq:objective2}--\eqref{eq:assignment2} are equivalent. The dual program \cite{balinski1969duality} of \eqref{eq:objective3} reads : 

\begin{align}
    \max & \sum_{k\in K} \mu_k   & \tag{D}\label{eq:dualobjective}\\
    \mbox{s.t. }    & \sum_{k\in K}a^k_I\mu_k\leq \Gamma_I& (\forall I\in \Lambda) \tag{D-1}\label{eq:dualconstraint}\\
    & \mu_k \geq 0 & (\forall k\in K)\tag{D-2}\label{eq:dualpositivity}
\end{align}

\vspace{5mm}
\noindent\textbf{Restriction and generation}
\vspace{5mm}

Let $\Lambda' \subset \Lambda$ be an arbitrary subset of allocations. One can form the primal dual pair of problems:

\begin{align} \tag{RCP}
    \min & \sum_{I \in \Lambda'} \Gamma_{I} x_{I} +R\sum_{k\in K} r_k   & \label{eq:objective3R}\\
    \mbox{s.t. }    & \sum_{I \in \Lambda'} a_I^kx_I + r_k \geq 1 & (\forall k\in K) \tag{RCP-1}\label{eq:assignment3R}\\
    & x_I, r_k \geq 0 & (\forall I\in \Lambda')(\forall k\in K)\tag{RCP-2}\label{eq:positivityR}
\end{align}
\begin{align}
    \max & \sum_{k\in K} \mu_k   & \tag{RD}\label{eq:dualobjectiveR}\\
    \mbox{s.t. }    & \sum_{k\in K}a^k_I\mu_k\leq \Gamma_I& (\forall I\in \Lambda') \tag{RD-1}\label{eq:dualconstraintR}\\
    & \mu_k \geq 0 & (\forall k\in K)\tag{RD-2}\label{eq:dualpositivityR}
\end{align}

Let $F$ (resp. $F'$) denote the feasible set of \eqref{eq:objective3} (resp. \eqref{eq:objective3R}) and $D$ (resp. $D'$) be the feasible set of \eqref{eq:dualobjective} (resp. \eqref{eq:dualobjectiveR}). It is clear that $F' \subset F$ and $D \subset D'$. Furthermore, for any $\Lambda' \subset \Lambda$, \eqref{eq:objective3R} is linear, feasible and bounded.
Therefore, strong-duality applies \cite{balinski1969duality}\cite{gale1951linear}. This means that \eqref{eq:objective3R} and \eqref{eq:dualobjectiveR} are both feasible and a primal-dual pair of solutions $(x',\mu')$ exists, where $x'$ solves \eqref{eq:objective3R}, $\mu'$ solves \eqref{eq:dualobjectiveR}. Furthermore, it means we have the equality:

\begin{equation}
    \sum_{I \in \Lambda'} \Gamma_{I} x'_{I} +R\sum_{k\in K} r'_k  = \sum_{k\in K} \mu'_k.
\end{equation}
The column generation is based on the following fact:

\begin{fact}
    Let $\Lambda' \subset \Lambda$. Let $(x^*,\mu^*)$ and $(x',\mu')$ be primal-dual \textit{optimal} couples for \eqref{eq:objective3}--\eqref{eq:dualobjective} and \eqref{eq:objective3R}--\eqref{eq:dualobjectiveR} respectively. By definition:
    \begin{itemize}
        \item $x^* \in F$
        \item $x' \in F' \subset F$
        \item $\mu^* \in D \subset D'$
        \item $\mu' \in D'$
    \end{itemize}
    
    Suppose that $\mu' \in D$. 

    Then, $x' \in F$ and $x'$ is optimal for \eqref{eq:objective3}. 
\end{fact}

Indeed, if $\mu' \in D$, by definition of $\mu^*$, we know that $\sum_{k\in K} \mu'_k \leq \sum_{k\in K} \mu^*_k$. On the other hand, as $D \subset D'$, by definition of $\mu'$, we have $\sum_{k\in K} \mu'_k \geq \sum_{k\in K} \mu^*_k$. Thus, we have equality. By strong duality, this means that 
\begin{align}
    \sum_{I \in \Lambda'} \Gamma_{I} x'_{I} +R\sum_{k\in K} r'_k &= \sum_{k\in K} \mu'_k \\
    & =  \sum_{k\in K} \mu^*_k\\
    & = \sum_{I \in \Lambda'} \Gamma_{I} x^*_{I} +R\sum_{k\in K} r^*_k 
\end{align}

Therefore, if $\mu'$ is feasible in \eqref{eq:dualobjective}, then $x'$ is an optimal solution of \eqref{eq:objective3}. 

For $\mu'$ to be feasible in \eqref{eq:dualobjective}, the following constraint must hold:
\begin{equation}\label{const}
\sum\limits_{k}a^k_I\mu'_k\leq \Gamma_I\text{ }\forall\text{ }I\in \Lambda
\end{equation}

Therefore, the existence of violated constraints \eqref{const} means that the current $x'$ is not the optimum and that new columns (allocations) can be added to improve the solution. We can therefore try to find an independent set that minimizes the \textit{reduced cost} $\Gamma_I-\sum\limits_{k}y^k_I\mu'_k$. If this reduced cost is negative then the independent set $I$ used to obtain this negative reduced cost violates \eqref{const} and can therefore be added to the set of independent sets in RCP $\Lambda'$.

For an allocation $I$, we have:
\begin{equation}
    \Gamma_I=\sum_k a_I^k\Gamma_{c_I}^k
\end{equation}

Therefore the reduced cost to minimize is:

\begin{equation}
\begin{split}
\sigma&=\Gamma_I-\sum\limits_{k}a^k_I\mu'_k \\
 &=\sum\limits_k a_I^k\Gamma_c^k-\sum\limits_{k}a^k_I\mu'_k \\
 &=\sum\limits_k a_I^k(\Gamma_c^k-\mu'_k)
\end{split}
\end{equation}

In order to convert this into a maximization problem, we can simply change the sign. The problem therefore reads:

\begin{equation}\label{mwiseqn}
\max \sigma = \sum\limits_k a_I^k(\mu'_k-\Gamma_c^k)
\end{equation}

By definition, $(a_I^k)_{k \in K}$ is a vector denoting an independent set, and $\mu'_k-\Gamma_c^k$ can be seen as numerical weights for every color $c$. This is therefore a Maximum Weighted Independent Set problem. For the rest of the paper, this problem will be our worker problem (WP). The MWIS problem can be defined as follows.

\begin{align}
    \max \sigma= & \sum_{k} y_k(\mu'_k-\Gamma_c^k)  & \tag{WP}\label{eq:sp}\\
    \mbox{s.t. }    & y_k+y_j \leq 1 &  (\forall (k,j)\in E)  \tag{WP-1}\label{eq:sp1}\\
    & y_k \in \{0,1\} & (\forall k \in K) \tag{WP-2}
\end{align}
 
A solution of Problem~\eqref{eq:sp} defines an allocation $(a_I, c)$ where $I \coloneqq \{k \in K | y_k = 1 \}$ and $a_I^k = y_k$. Note that there is one worker problem per color, and, as in our problem, colors from the same model are equivalent, there is actually one problem per vehicle model. Therefore, each allocation produced by a worker represents a single physical vehicle (color) along with all the tours assigned to it. Given the above problem definitions, we can define the algorithm to solve RCP as described in Algorithm~\ref{alg:one}.

\begin{algorithm}[htb!]
\caption{Column Generation algorithm for RCP}\label{alg:one}
\DontPrintSemicolon
\KwIn{Incompatibility graph $G(K,E)$}
Define $R$ such that $R > \max \Gamma_I + 1$\;
$\Lambda' \gets \emptyset$\;
\While{True}{
Solve RCP and get primal dual couple $(x',\mu')$\;
\For{every model $v$} {Solve WP \\ \If {$\sigma>0$} {Let the obtained solution be $I=(y_I,c_I)$.\\ $\Lambda'\gets \Lambda' \cup \{I\}$.}} \;
\If{$\Lambda'$ has been modified} {continue} \Else{Current solution $x'$ is the optimal solution \\ break}} \;

Apply rounding algorithm to transform the relaxed solution $x'$ into a binary one.


\end{algorithm}

\subsection{Quantum model for the MWIS problem}\label{QuantumModel}

The MWIS problem can be represented in the form of a Quadratic Unconstrained Binary Optimization (QUBO) problem. Our problem in question is the problem WP. In \eqref{eq:sp}, for simplicity, let  $\mu'_k-\Gamma_c^k=w_k$. The objective function is therefore $\sum\limits_k y_k w_k$. The constraint \eqref{eq:sp1} can be incorporated into the objective function as a penalty \cite{qubotutorial} as follows. The real number $P$ is the penalty strength. Note that the variable $y$ in this section is not related to the other variables $y$ that have appeared previously.

\begin{equation}\label{qubo_primary}
    \mathcal{C}=\sum\limits_k y_k w_k - P\bigg(\sum\limits_{(i,j)\in E} y_i y_j\bigg)
\end{equation}

Since, $y \in \{0,1\}$, we can set $y_k^2=y_k$. Therefore:

\begin{equation}\label{qubo_original}
    \mathcal{C}=\sum\limits_k y_k^2 w_k - P\bigg(\sum\limits_{(i,j)\in E} y_i y_j\bigg)
\end{equation}

Since $\mathcal{C}$ is in the form of a QUBO problem, we can represent it in the following form:

\begin{equation}\label{quboeqn}
    \mathcal{C}=y^T\mathcal{Q}y 
\end{equation}

where $y$ is the vector representation of the independent set and $\mathcal{Q}$ is the QUBO matrix.

As described in Algorithm 3 of \cite{chatterjee2023solving}, a QUBO problem can be represented on a quantum computer using only a logarithmic number of qubits. We shall use this algorithm to solve WP. Following are some of the main aspects of the algorithm. Our main aim here is to represent \eqref{quboeqn} on a quantum computer.

\begin{enumerate}

\item Let the size of the problem be $n$, then the number of qubits required will be $N=\lceil \log_2 n \rceil$. A $N$-qubit parameterized state $\ket{\Psi(\theta)}$ is created as follows:
\begin{equation}\label{statevector}
    \ket{\Psi(\theta)}= U(\theta)H^{\otimes N}\ket{0}^{\otimes N}
\end{equation}

The equation above represents the application of $N$ Hadamard ($H$) gates on $N$ qubits, all initially in state $\ket{0}$; followed by the application of a diagonal gate $U(\theta)$ which is of the following form:
\begin{equation}\label{U_theta}
         U(\theta)=  \begin{bmatrix}
                e^{i \pi R(\theta_1)} & 0 & 0 & ....\\
                0 & e^{i \pi R(\theta_2)} & 0 & ....\\
                ....&....& ....&....\\
                0&0&0&e^{i \pi R(\theta_n)}
            \end{bmatrix}
\end{equation}

where \begin{equation}\label{R_theta}
        R(\theta_k)=\begin{cases}
          0\quad &\text{if } \, 0 \leq \theta_k <\pi \\
          1 \quad &\text{if } \, \pi \leq \theta_k < 2\pi \\
     \end{cases}
    \end{equation}

Using \eqref{U_theta} and \eqref{R_theta} in \eqref{statevector},  we have:
\begin{equation}\label{psi_theta}
         \ket{\Psi(\theta)}=  \begin{bmatrix}
                e^{i \pi R(\theta_1)} \\
                e^{i \pi R(\theta_2)}\\
                ......\\
                e^{i \pi R(\theta_n)}
            \end{bmatrix}
\end{equation}

$\ket{\Psi(\theta)}$ is therefore a vector whose terms belong to the set $\{1,-1\}$.

\item Note that when we defined the QUBO in equation \eqref{quboeqn}, the variables were $y\in \{0,1\}$. Since $\ket{\Psi(\theta)}\in \{1,-1\}$, we need to define our QUBO matrix using variables $z \in \{1,-1\}$. This new matrix is then called the spin-QUBO or sQUBO as defined in \cite{chatterjee2023solving}. We shall denote the sQUBO matrix as $\mathcal{Q}'$, which is a matrix of size $2n\times 2n$. For details regarding how to generate $\mathcal{Q}'$ see Appendix \ref{sQUBO}.

    \item Equation \eqref{quboeqn} can have the following quantum equivalent:
     \begin{equation}\label{expval}
    \mathcal{C}(\theta)=-\bra{\Psi(\theta)}\mathcal{Q}'\ket{\Psi(\theta)} + K
\end{equation}

where $\theta=\{\theta_1...\theta_n\}$ is a set of parameters and $\ket{\Psi(\theta)}$ is a parameterized ansatz that represents that vector $y$ and $K$ is a constant. The interesting thing about this representation is that we need only 
 $1+\log n$ qubits to represent $\mathcal{Q}'$ of size $2n\times 2n$. Therefore, only $N=1+\lceil \log n \rceil$ qubits will be required if we have a problem of size $n$. We put the negative sign as we want to make it a minimization problem.

\item Measure the expectation value \eqref{expval} on a quantum computer or a simulator. In order to calculate this expectation value we need to decompose $\mathcal{Q}'$ into a sum of Pauli strings (see Appendix \ref{expectation_value}).

\item Using a classical optimizer such as the genetic algorithm (GA), optimize the parameters $\theta$. 

\begin{equation}
  \mathcal{C}^*(\theta^*)=\min \mathcal{C}(\theta) =- \sigma
\end{equation}

From the optimized parameters we can get the binary solution using:

\begin{equation}\label{thetatobinary}
       y_k= \frac{1-\exp (i\pi R(\theta_k^*))}{2}=\begin{cases}
          0\quad &\text{if } \, 0 \leq \theta_k^* <\pi \\
          1 \quad &\text{if } \, \pi \leq \theta_k^* < 2\pi \\
     \end{cases}
    \end{equation}
\end{enumerate}

\subsection{Quantum-assisted algorithm to solve the Coordinator Problem}\label{QuantumAlg}

We now have a quantum model to solve the WP. We will call this the Quantum Worker Solver(QWS). To develop a quantum-assisted algorithm, the QWS is used together with a classical worker solver (CWS) to reach the optimal solution. The aim is to use the QWS together with the CWS where the CWS is used \textit{only} when the QWS is not able to generate a new column.

\begin{figure}[htb!]
  \includegraphics[width=\linewidth]{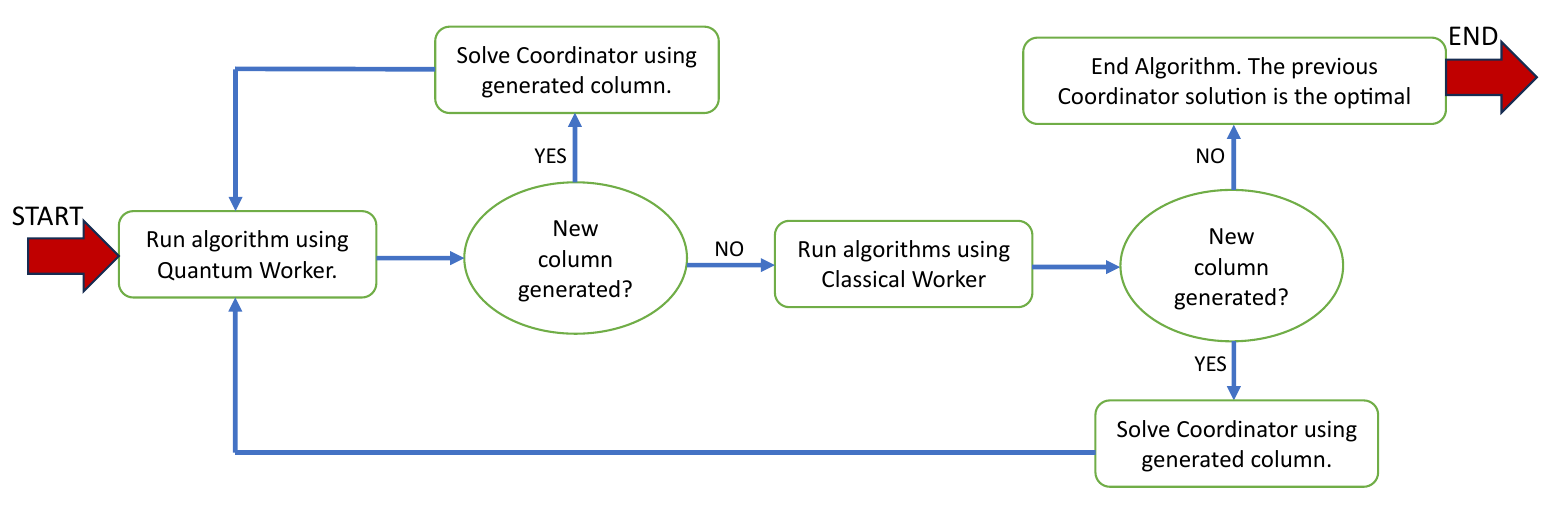}
  \caption{Progress of the quantum-assisted Algorithm}
  \label{algo}
\end{figure}

For every instance, we start the algorithm with the QWS as the worker. If the algorithm is able to generate a new column we use it to solve the coordinator (RCP) and continue with the optimization loop using the QWS. Note that generating a new column here means that the MWIS solution found gives a positive $\sigma$ (as in equation \eqref{mwiseqn}), hence violating \eqref{const}. In case QWS is not able to generate a new column, we check whether the CWS can generate a new column. If the CWS is able the generate a new column, we solve the RCP using this new column and continue the optimization loop with the QWS. If even the CWS is unable to generate a new column, we have the optimal solution to the problem as the previous solution to the RCP obtained. This is explained diagrammatically in figure \ref{algo}.

Using figure \ref{algo}, we can modify Algorithm~\ref{alg:one} to get Algorithm~\ref{alg:two}.

\begin{algorithm}[htb!]
\caption{Quantum-assisted Column Generation algorithm for RCP}\label{alg:two}
\DontPrintSemicolon
\KwIn{Incompatibility graph $G(K,E)$}
Define $R$ such that $R > \max \Gamma_I + 1$\;
$\Lambda' \gets \emptyset$\;
\While{True}{
Solve RCP and get primal dual couple $(x',\mu')$\;
\For{every model $v$} {Solve WP using QWS\\ \If {$\sigma>0$} {Let the obtained solution be $I=(y_I,c_I)$.\\ $\Lambda'\gets \Lambda' \cup \{I\}$.}} \;
\If{$\Lambda'$ has been modified} {continue} \Else{\For{every model $v$} {Solve WP using CWS\\ \If {$\sigma>0$} {Let the obtained solution be $I=(y_I,c_I)$.\\ $\Lambda'\gets \Lambda' \cup \{I\}$.}} \If{$\Lambda'$ has been modified} {continue} \Else{Current solution $x'$ is the optimal solution \\ break}}} \;

Apply rounding algorithm to transform the relaxed solution $x'$ into a binary one.


\end{algorithm}

\section{Experimental Results}\label{Experiments}

In this section, we demonstrate experiments carried out using tours of sizes 32 and 64. Synthetic data was generated with start and end times of tours as well as sets of allowed vehicles for every tour. We decide to have five different vehicle types and the cardinality of the set of allowed vehicles is 3. For simplicity, we do not generate location data. It can, however be added easily which will only change the density of the incompatibility graph. All runs of the quantum algorithm in this section are done using a quantum simulator.

\begin{table}[htb!]
\begin{center}
  \begin{tabular}{|c|c|c|c|c|}
    \hline
   Instance Size & Number of Qubits & Number of Instances & Mean $\%$ of Quantum Iterations \\ 
   \hline
   32 & 6 & 5
 & 87.36\\
   64 & 7 & 5 & 81.73\\
\hline
  \end{tabular}
  \caption{\centering Results for 32 and 64 Tours.}
   \label{ResultsTable}
   \end{center}
\end{table}

\begin{table*}[htb!]
\begin{center}
  \begin{tabular}{|l|l|l|}
    \hline
   \multirow{2}{*}{} & \multicolumn{2}{c|}{Instance Size} \\ 
   \cline{2-3} 
    & \multicolumn{1}{c|}{32} & \multicolumn{1}{c|}{64} \\
   \hline
   Solver used for RCP & \multicolumn{2}{c|}{Gurobi}\\
   \hline
   Solver used for CWS & \multicolumn{2}{c|}{Gurobi}\\
   \hline
   Classical optimizer used for variational loop of QWS & \multicolumn{2}{c|}{Genetic algorithm (python package: geneticalgorithm)} \\
   \hline
   Penalty value & \multicolumn{1}{c|}{10} & \multicolumn{1}{c|}{20} \\
   \hline
 
     \multicolumn{1}{|p{6cm}|}{Average time taken to simulate a complete run of the algorithm on an instance} & \multicolumn{1}{p{3cm}|}{$\approx 70$ minutes} & \multicolumn{1}{c|}{ $\approx 24$ hours} \\
\hline
  \end{tabular}
  \caption{\centering Supplementary information regarding the experiments}
   \label{Supplementary}
   \end{center}
\end{table*}

\begin{table*}[htb!]
\begin{center}
  \begin{tabular}{|l|l|l|}
    \hline
   \multirow{2}{*}{Genetic Algorithm Parameter} & \multicolumn{2}{c|}{Instance Size} \\ 
   \cline{2-3} 
    & 32 & 64 \\
   \hline
   max\_num\_iteration & 50 & 100\\
   \hline
   population\_size & 20 & 40\\
   \hline
   mutation\_probability & \multicolumn{2}{c|}{0.1} \\
   \hline
   elit\_ratio & \multicolumn{2}{c|}{0.05} \\
   \hline
   crossover\_probability & \multicolumn{2}{c|}{0.5}\\
   \hline
   parents\_portion & \multicolumn{2}{c|}{0.3}\\
   \hline
   crossover\_type & \multicolumn{2}{c|}{uniform}\\
   \hline
   max\_iteration\_without\_improv & \multicolumn{2}{c|}{None} \\
   
\hline
  \end{tabular}
  \caption{\centering Parameters of the Genetic Algorithm used in the experiments}
   \label{Supplementary2}
   \end{center}
\end{table*}

\begin{table}[htb!]
\begin{center}
  \begin{tabular}{|c|c|c|c|c|}
    \hline
   Instance Size & No. of Qubits & GA Population Size & GA Iterations for QWS & Estimated Time (hours) \\ 
   \hline
   128 & 8 & 20 & 100
 & 108\\
   256 & 9 & 20 &100 & 3500\\
\hline
  \end{tabular}
  \caption{\centering Estimated time for simulating instances of 128 and 256 Tours.}
   \label{TimeEstimate}
   \end{center}
\end{table}

In order to evaluate the contribution of the QWS, the number of successful quantum solves is compared to the total number of successful solves. A successful solve means that the QWS or CWS was able to generate a new column. In table \ref{ResultsTable}, the number of quantum solves is shown as a percentage of the total number of solves, averaged over all the instances. 

Tables \ref{Supplementary} and \ref{Supplementary2} give some important details regarding the experiments.

\begin{figure}[htb!]
  \includegraphics[width=\linewidth]{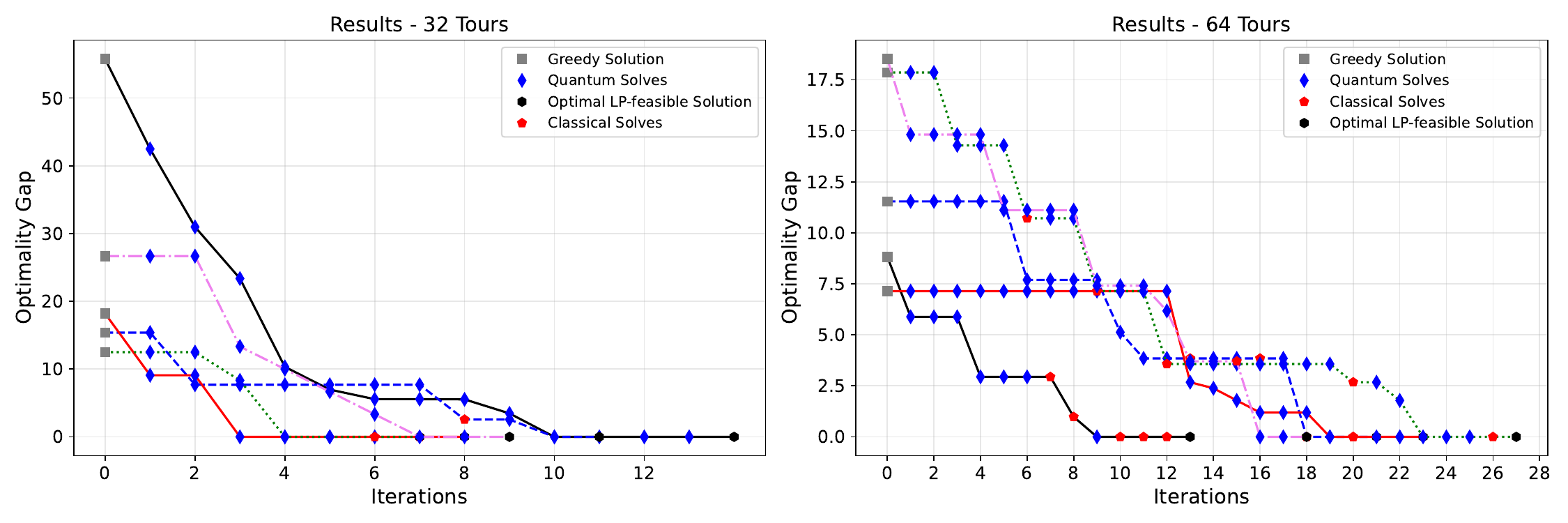}
  \caption{Full run of the algorithm for instances of 32 and 64 tours. }
  \label{128results}
\end{figure}

Figure \ref{128results} shows in detail complete runs of algorithm \ref{alg:two} for instances of size 32 and 64. For every run, the first point of a plot is a gray square. This is a quick greedy solution to the problem and is the starting point of our algorithm. This greedy algorithm colors each node one by one, ensuring that neighboring nodes do not share the same color. This ensures that we already have a decent solution to start with. As we go forward in the plots, we have blue diamonds and red pentagons which signify the RCP solutions obtained using the QWS and the CWS respectively. Finally we have the black hexagon which is the optimal point, where both the QWS and CWS were not able generate a new column. The cost function has been normalized using the formula : $\frac{\text{(Current Solution)}-\text{(Optimal LP-feasible Solution)}}{\text{Optimal LP-feasible Solution}}$.

The iterations of the algorithm are different from and not to be confused with the GA iterations of the QWS.

Taking the instance with the blue dash-dotted line for 64 tours as an example, there are 16 iterations where the QWS was successful, 1 where QWS was unsuccessful and CWS was successful and 1 where both QWS and CWS were unsuccessful. Note, however, that for every iteration we called the QWS. Hence, there were a total of 18 QWS calls. Every QWS call requires 5 MWIS solutions (one for each vehicle type) each requiring around 300 expectation value measurements, hence the total number of expectation value measurements required were of the order of $10^4$. Even though the size was tractable to represent our problem, it was therefore not feasible to run the algorithm on a quantum computer since every expectation value measurement on an IBM quantum computer can take from a few seconds to a few hours, depending upon the wait time (or queue time). In addition we were limited to a size of 64 nodes for simulation since for 128 nodes and above, the simulation time was too long to carry out experiments. We can see in Table \ref{TimeEstimate} the estimated time that might be required to potentially simulate instances of 128 and 256 tours. These estimates are given considering the population-size and number of GA iterations to be 20 and 100 respectively, and considering that there are 25 algorithm iterations to converge to a solution. Here, the QPU should have significant a time advantage over the simulators. 

\begin{figure}[htb!]
\centering
  \includegraphics[width=0.8\linewidth]{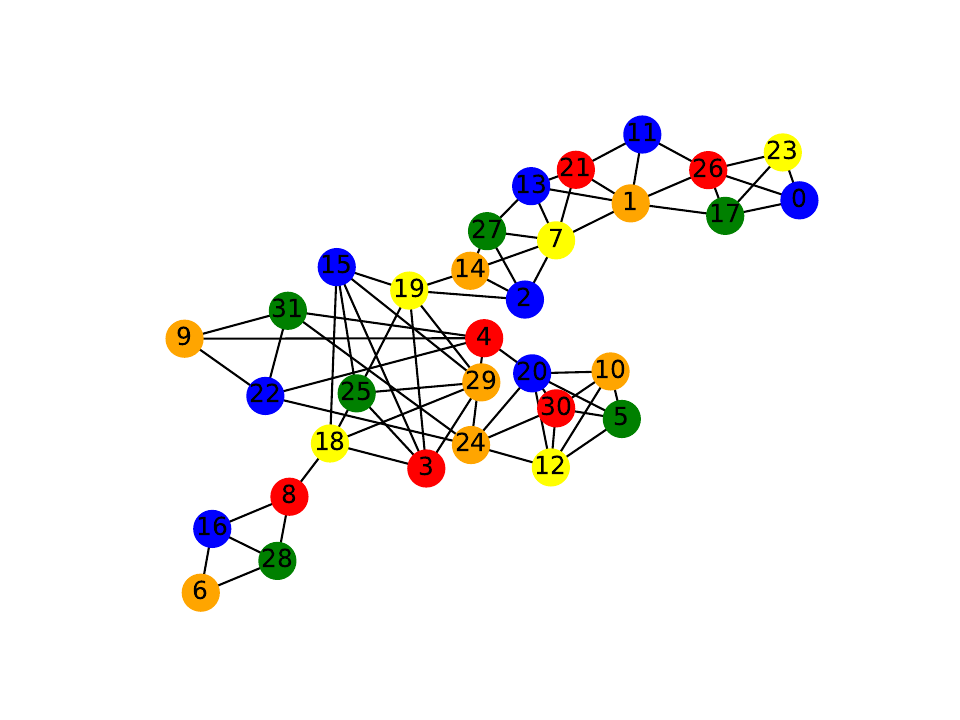}
  \caption{Incompatibility graph for a 32-tour instance.}
  \label{draw_instance}
\end{figure}

\begin{figure}[htb!]
  \includegraphics[width=\linewidth]{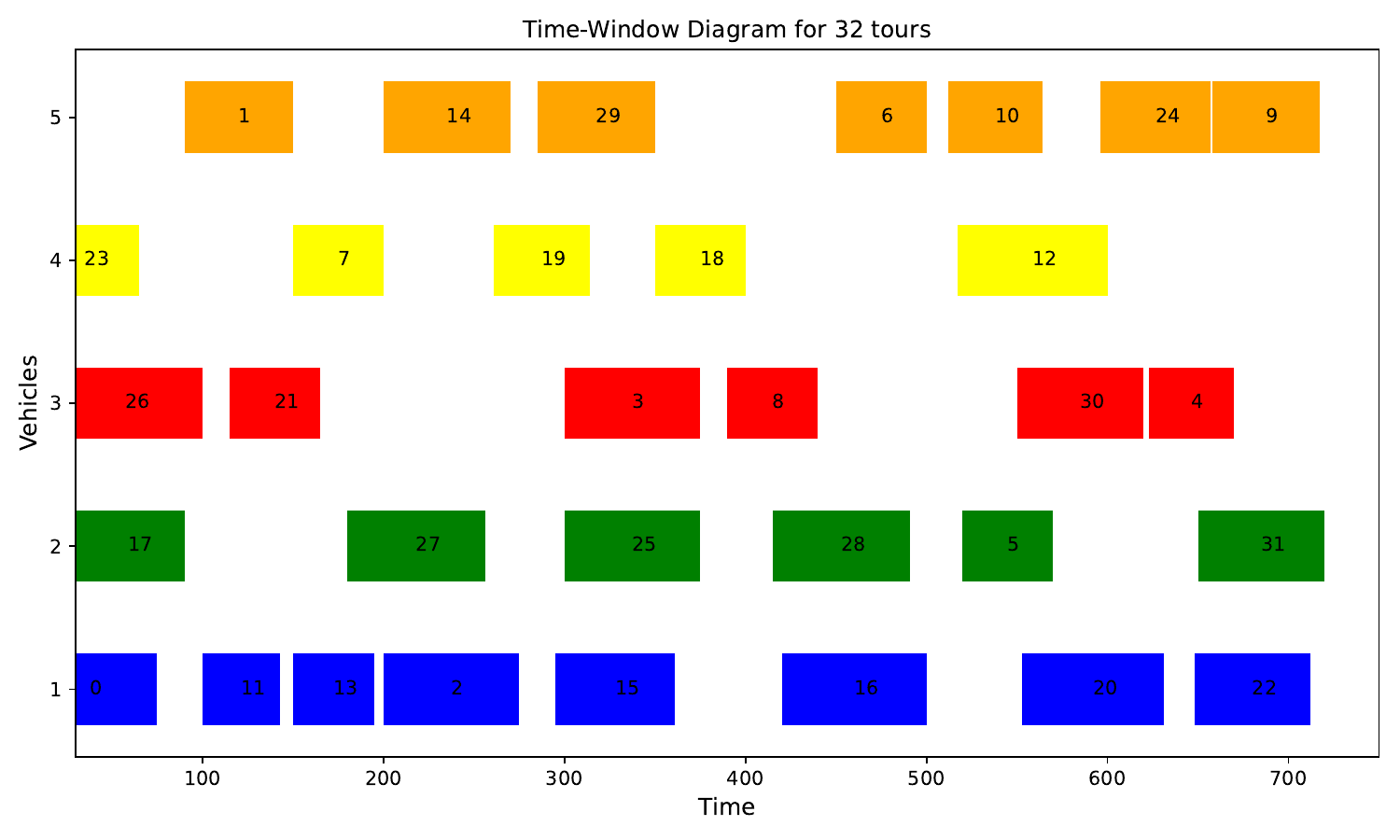}
  \caption{Time-window diagram for a 32-tour instance.}
  \label{draw_solution}
\end{figure}

Finally, figures \ref{draw_instance} and \ref{draw_solution}  show the result of a 32-tour instance. This is specifically the instance denoted by a black solid line in figure \ref{128results}. Figure \ref{draw_instance} shows the incompatibility graph and figure \ref{draw_solution} shows the corresponding time-window diagram. The nodes in figure \ref{draw_instance} do not refer of any specific location. The tour numbers from 0 to 31 are marked on the time-window bars as well as the nodes. The 5 vehicles used to carry out all the tours are marked using different colors. 

\section{Discussion and Conclusion} 

In this paper we have successfully demonstrated a quantum-assisted algorithm to solve the Fleet Conversion problem. This shows firstly the advantage of having an algorithm that scales logarithmically with the size of the problem. This trait helped us model problems of sizes that are well outside the realm of what other quantum variational algorithms can handle. Notably, our algorithm successfully handles instances that, albeit synthetic, represent realistic, non-trivial industrial problem sizes, which displays the potential utility of leveraging quantum computing for solving complex optimization problems.

Secondly, it demonstrates  the possibility of harnessing quantum algorithms in conjunction with classical solvers rather than being in competition with them. If the problem is very resource intensive then instead of the using a full classical optimization, a part of the computation can be transferred to the quantum computer, potentially reducing required computational resources (logarithmic representation). 

There remain, however, several drawbacks in the algorithm. Firstly, the algorithm requires the decomposition of the Hamiltonian matrix into a sum of Pauli strings. The cost of doing this using the method used here (as shown in appendix \ref{expectation_value}) increases significantly with an increase in the number of qubits. Reducing the time to decompose the matrix could reduce runtime by a large margin for larger instances.

Secondly, larger instances require a larger number of runs on the genetic algorithm and therefore it quickly becomes infeasible to increase problem size. Even if we decide not to increase the number of genetic algorithm runs, the quality of the QWS solution will degrade resulting in more QWS calls, hence a higher number of iterations (considering we require the same level of performance). 

The success of the algorithm depends on the accuracy of the QWS solution. Worse QWS solutions inadvertently lead to more QWS calls. Hence the runtime of the algorithm could be quite sensitive to noise in potential runs on a quantum computer. The indicates that this algorithm would be suitable for small but fault tolerant quantum computers rather than larger ones with higher levels of noise. Despite the fact that we are moving towards a large number of physical qubits in the future, the availability of logical (fault-tolerant) qubits will remain limited. Consequently, this approach will continue to be relevant.

The availability and efficiency of quantum hardware remain subjects of ongoing research. Despite this uncertainty, notable advancements in the quantum hardware field in recent years, coupled with the qubit-efficiency of the demonstrated algorithm suggests that quantum-assisted algorithms may hold significant promise in the future.

\section*{acknowledgments}
Y.C. and M.R. acknowledge funding from European Union’s Horizon 2020 research and innovation programme, more
specifically the $\bra{NE}AS\ket{QC}$ project under grant agreement No. 951821.  A part of the methodology used in the
manuscript is protected by a provisional patent claim ”Method for optimizing a functioning relative to a set of elements
and associated computer program product” submission number EP21306155.9 submitted on 26.8.2021.

\section*{conflict of interest}
The authors declare no conflict of interest.

\printendnotes

\bibliography{main}

\appendix

\section{Converting QUBO to spin-QUBO} \label{sQUBO}

Let $z\in \{1,-1\}$ and $y\in \{0,1\}$, then it is quite easy to verify that $y=\dfrac{1-z}{2}$. Using this in equation \eqref{qubo_primary}, we have:

\begin{equation}\label{qubonew}
\mathcal{C}=\sum\limits_k  w_k \dfrac{1-z_k}{2} - P\bigg(\sum\limits_{(i,j)\in E} \dfrac{1-z_i}{2} \dfrac{1-z_j}{2}\bigg)
\end{equation}

In the search for optimal values of parameters $z$ we can eliminate the constant terms in $\mathcal{C}$ as they only add a constant shift to the cost function. We can therefore aggregate the constant terms to an overall constant $K$. We can therefore simplify \eqref{qubonew} as follows:

\begin{align}
\mathcal{C}&=\sum\limits_k  w_k \dfrac{1-z_k}{2} - P\bigg(\sum\limits_{(i,j)\in E} \dfrac{1-z_i}{2} \dfrac{1-z_j}{2}\bigg)\\
       &=\dfrac{1}{2}\sum_{k} w_{k} (1-z_k) - \dfrac{P}{4}\sum_{\substack{ij \\ i \neq j}}  (1-z_i-z_j+z_iz_j)\\
       &=-\dfrac{1}{2}\sum_{k} w_{k} z_k + \dfrac{P}{4}\sum_{\substack{ij \\ i \neq j}} (z_i+z_j-z_iz_j)+K\label{qubonew1} 
\end{align}

The above \textit{cannot} be represented as a QUBO matrix since it has linear terms which cannot be quadratized since $z_i^2 \neq z_i$.

Hence, we need to reformulate the problem. In this reformulation the linear terms are represented in the off-diagonal terms instead of the diagonals. Let us have $2n$ variables $\{z_1,z_2....z_{2n}\}$ where $z_1... z_n \in \{1,-1\}$ and $z_{n+1}... z_{2n} \in \{1\}$. Then to represent a linear variable $z_i$, we can have the term $z_iz_{i+n}$ where $z_{i+n}=1$. The equation \eqref{qubonew1} can be rewritten as follows:

\begin{equation}\label{qubonew2}
    \mathcal{C}=-\dfrac{1}{2}\sum_{k} w_k z_kz_{k+n} + \dfrac{P}{4}\sum_{\substack{ij \\ i \neq j}} (z_iz_{i+n}+z_jz_{j+n}-z_iz_j) +K
\end{equation}

This can now be written as:
\begin{equation}\label{sQUBO_final_format}
   \mathcal{C}=z^T\mathcal{Q}'z + K
\end{equation}

This is the required $\mathcal{Q}'$. It is a matrix of size $2n\times 2n$.

\section{Calculating the Expectation value of an observable} \label{expectation_value}

Given the observable $\mathcal{O}$, we first need to convert into a sum of tensor products of Pauli strings. 

Let $\mathcal{O}$ be a $n\times n$ matrix and $S =\{I,X,Y,Z\}^N=\{S_1,S_2,S_3,S_4\}^N $ be the set of Pauli matrices. We can consider $n$ to be a power of $2$ without any loss of generality. If the size of the Hamiltonian matrix is $n'$ which is not a power of $2$, we can easily convert it to a size of $n=2^{\log_2(n')}$, which is a power of $2$. The extra space in the matrix is filled with $0$'s.

This Hamiltonian can now represented on $N=\log_2(n)$ qubits. Consider the set $J=\{S_{i_1}\otimes S_{i2}...\otimes S_{iN}| i_1,i_2....i_N \in \{0,1,2,3\}  \}$ which consists of all tensor product combinations of the Pauli matrices.

Then the Hamiltonian can decomposed as:
 
\begin{equation}
    \mathcal{O}=\sum\limits_{i=1}^{4^N} c_iJ_i
\end{equation}
where the coefficients are:
\begin{equation}
    c_{i}=\frac{1}{n}Tr(J_i\cdot \mathcal{O})
\end{equation}

The Hamiltonian therefore becomes:

\begin{equation}
    \mathcal{O}=\frac{1}{n}\sum\limits_{i=1}^{4^N}Tr(J_i\cdot \mathcal{O})J_i
\end{equation}

The  expectation value becomes a sum of the expectation values of all the terms.

\begin{equation}
    \bra{\Psi}\mathcal{O}\ket{\Psi}= \frac{1}{n}\sum\limits_{i=1}^{4^N}Tr(J_i\cdot \mathcal{O})\bra{\Psi}J_i\ket{\Psi}
\end{equation}
\end{document}